\documentclass[%
 reprint,
superscriptaddress,
 amsmath,amssymb,
 aps,
prb,
]{revtex4-2}

\usepackage{graphicx}
\usepackage{dcolumn}
\usepackage{bm}
\usepackage{braket}
\usepackage{hyperref,xcolor}
\hypersetup{
	colorlinks,
	linkcolor={magenta},
	citecolor={magenta},
	urlcolor={blue!70!black}
}

\begin{document}


\title{Quantum Hall effect induced by chiral Landau levels in topological semimetal films}

\author{D.-H.-Minh Nguyen}
\email{d.h.minh.ng@gmail.com}
\affiliation{Institute for Materials Research, Tohoku University, Sendai, Miyagi, 980-8577, Japan}
\author{Koji Kobayashi}
\affiliation{Institute for Materials Research, Tohoku University, Sendai, Miyagi, 980-8577, Japan}
\author{Jan-Erik R. Wichmann}
\affiliation{Institute for Materials Research, Tohoku University, Sendai, Miyagi, 980-8577, Japan}
\author{Kentaro Nomura}
\email{nomura@imr.tohoku.ac.jp}
\affiliation{Institute for Materials Research, Tohoku University, Sendai, Miyagi, 980-8577, Japan}%
\affiliation{Center for Spintronics Research Network, Tohoku University, Sendai 980-8577, Japan}

\date{\today}

\begin{abstract}
Motivated by recent transport experiments, we theoretically study the quantum Hall effect in topological semimetal films. Owing to the confinement effect, the bulk subbands originating from the chiral Landau levels establish energy gaps that have quantized Hall conductance and can be observed in relatively thick films. We find that the quantum Hall state is strongly anisotropic for different confinement directions not only due to the presence of the surface states but also because of the bulk chiral Landau levels. As a result, we re-examine the quantum Hall effect from the surface Fermi arcs and chiral modes in Weyl semimetals and give a more general view into this problem. Besides, we also find that when a topological Dirac semimetal is confined in its rotational symmetry axis, it hosts both quantum Hall and quantum spin Hall states, in which the helical edge states are protected by the conservation of the spin-$z$ component.
\end{abstract}

\maketitle


\section{Introduction}
Weyl orbits \cite{Potter2014,Zhang2016} in topological semimetals are unique magnetic orbits that involve two Fermi arcs on opposite surfaces connected via the chiral Landau levels running through the bulk. Experiments studying the quantum oscillation and quantum Hall effect (QHE) related to such orbits have been carried out in recent years \cite{Zhang2019a,Zhang2017,Nakazawa2019,Nishihaya2019,Goyal2018,Galletti2019,Schumann2018,Lin2019,Galletti2018a,Zheng2017,Moll2016} and have sparked heated debates among research groups. For instance, as most of the experiments were conducted with the topological Dirac semimetal (DSM) candidate Cd$_3$As$_2$ \cite{Wang2013,Borisenko2014,Liu2014,He2014,Neupane2014}, some works attribute the QHE in those Cd$_3$As$_2$ films to the surface Dirac cones instead of the Weyl orbits \cite{Galletti2018a,Kealhofer2020}. This is because the surface states of Cd$_3$As$_2$ are shown to be two-dimensional (2D) Dirac cones by the angle-resolved photoemission measurements \cite{Yi2014a,Neupane2015,Roth2018a} instead of the open Fermi arcs as theoretically predicted \cite{Wang2013}. In order to distinguish the two mechanisms of QHE in Cd$_3$As$_2$, it is proposed as a key that the QHE based on Weyl orbits depends on the film thickness \cite{Zhang2019a,Goyal2018,Galletti2019,Zhang2016}, which was observed in a wedge-shaped film of Cd$_3$As$_2$ \cite{Zhang2019a}. However, a recent study suggests that the QHE in such wedge-shaped films may come from a completely different mechanism \cite{Cheng2020b}, and thus the existence of Weyl orbits remains elusive. As another example for the debates, some experiments \cite{Zhang2019a,Zhang2017} find that the energy levels of Weyl orbits are related to the quantum confinement subbands, which contradicts with some other works \cite{Nishihaya2019,Wang2017,Li2020b} distinguishing Weyl orbits with the confinement effect. Hence, further studies are still needed in order to understand the nature of both Weyl orbits and the surface states of Cd$_3$As$_2$.

Motivated by the connection between Weyl orbits and the quantum confinement effect, and by the progress in fabricating high-quality nanostructures of Cd$_3$As$_2$ \cite{Kealhofer2019,Galletti2018}, we investigate the QHE induced by confinement subbands of the chiral Landau levels in topological semimetal films. We study the QHE in Weyl semimetals (WSMs) in two cases of the confinement direction: with and without the surface states. While in the former, the energy levels agree with the results obtained from the semiclassical analysis as expected, we show that the chiral Landau subbands behave differently in the latter. By examining the difference between those two cases, we find a more general expression for the energy levels of Weyl orbits in a WSM. Furthermore, we find that in DSM films confined in their rotational symmetry axis, the quantum Hall and quantum spin Hall states coexist due to the spin conservation.

This paper is organized as follows. In Sec. \ref{sec2}, we introduce the models describing our WSM and DSM. In Sec. \ref{sec3}, we revisit the QHE based on Weyl orbits by considering a WSM film with chiral surface states on their boundaries. We then carry out the same calculation but in a film confined in a direction so that it has no surface states. In Sec. \ref{sec4}, we explain the results obtained in the preceding section and examine the QHE in a WSM confined in an arbitrary direction. We also study the QHE in a DSM confined in its rotational symmetry axis and consider a specific case of $(001)$ Cd$_3$As$_2$ film to show that the effect is experimentally observable. Finally, our results will be summarized in Sec. \ref{sec5}.
\section{Models}\label{sec2}
\begin{figure}
	\includegraphics[width=\linewidth]{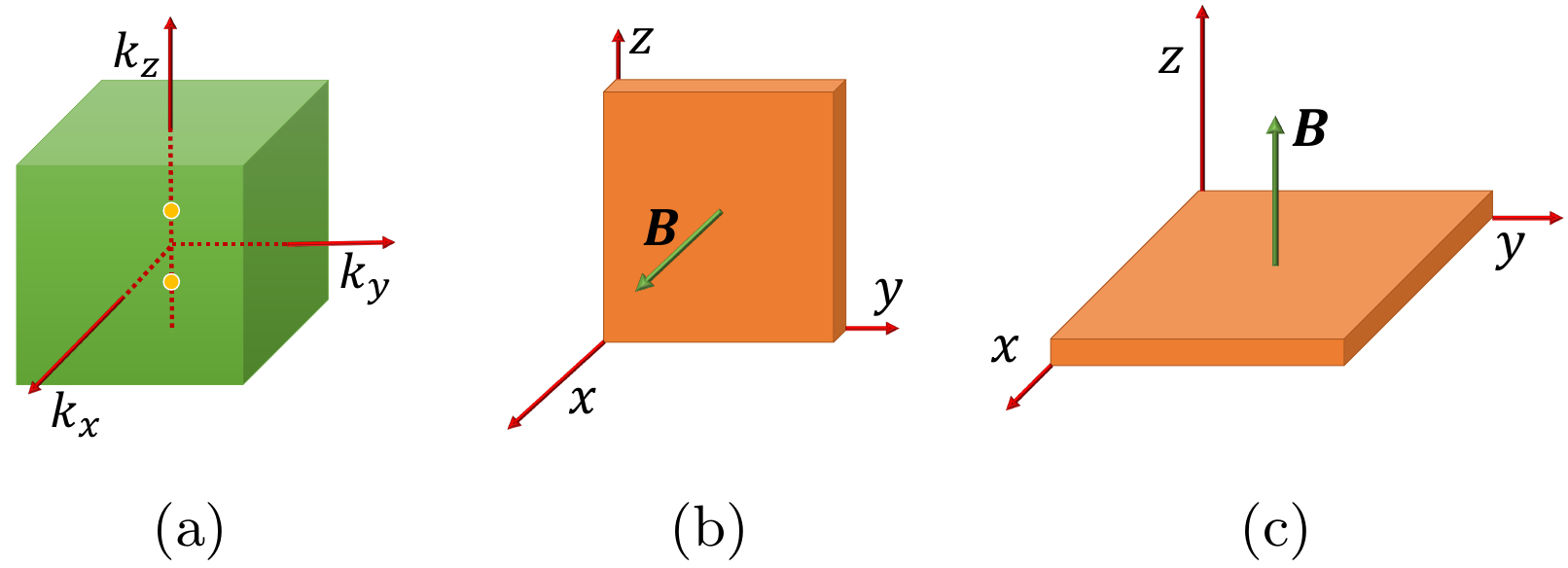}
	\caption{\label{fig:1} (a) WSM/DSM with two Weyl/Dirac nodes (orange points) aligned along the $z$ axis. (b)/(c) A slab confined in the $x$/$z$ direction in a perpendicular magnetic field $\mathbf{B}$.}
\end{figure}

In this work, we describe a WSM by the minimal model \cite{Yang2011,Imura2011}
\begin{align}
	H_{\text{W}}(\mathbf{k}) = & t(\sin k_x\sigma_x + \sin k_y\sigma_y) \nonumber\\
	&+ [M + t'(\cos k_x + \cos k_y + \cos k_z)]\sigma_z\label{Eq: Weyl}
\end{align}
with $\sigma_i$ being the Pauli matrices. Notice that the dimensionless $k_i$ denote the product $q_ia_i$, where $q_i$ are the usual crystal momenta and $a_i$ lattice constants. This cubic lattice model is centrosymmetric and breaks the time-reversal symmetry. Hereafter, we use $t=t'=1$ and $M=-2.5$ so that the Hamiltonian gives a pair of Weyl nodes at $\mathbf{k}_W=(0,0,\pm\pi/3)$, as depicted in Fig.~\hyperref[fig:1]{1(a)}. If we confine this WSM in the $x$ direction, the topological surface states called Fermi arcs will appear. In our minimal model, the arcs are straight lines connecting the projections of the two bulk Weyl cones.

When the Weyl points are located close to the $\Gamma$-point, we can expand the Hamiltonian to obtain a continuum model
\begin{equation}
	H_{\text{W}}(\mathbf{k}) = t(k_x\sigma_x + k_y\sigma_y) + [m - t'(k_x^2+k_y^2+k_z^2)/2]\sigma_z,\label{Eq: Weylconti}
\end{equation}
where $m=M+3t'$. 
We will use the lattice Hamiltonian for numerical calculations and its continuum counterpart for obtaining the analytical expressions.

On the other hand, the DSM is described by an effective Hamiltonian of Cd$_3$As$_2$ and Na$_3$Bi \cite{Wang2013,Wang2012a}
\begin{equation}
	H_{\text{D}}(\mathbf{q}) = \tilde{\mathcal{E}}_0(\mathbf{q}) + \begin{pmatrix}
		\tilde{\mathcal{M}}(\mathbf{q}) & Aq_+ & 0 & 0 \\
		Aq_- & -\tilde{\mathcal{M}}(\mathbf{q}) & 0 & 0\\
		0 & 0 & \tilde{\mathcal{M}}(\mathbf{q}) & -Aq_-\\
		0 & 0 & -Ak_+ & -\tilde{\mathcal{M}}(\mathbf{q})
	\end{pmatrix} \label{Eq: Diracconti}
\end{equation}
with $q_\pm = q_x\pm iq_y$, $\tilde{\mathcal{E}}(\mathbf{q}) = C_0 + C_1q_z^2 + C_2(q_x^2+q_y^2)$, and $\tilde{\mathcal{M}}(\mathbf{q}) = M_0 + M_1q_z^2 + M_2(q_x^2+q_y^2)$. We transform this into a tetragonal lattice model using the substitution
\begin{equation}
	q_i\rightarrow \frac{1}{a_i}\sin k_i,\qquad q_i^2\rightarrow \frac{2}{a_i^2}(1-\cos k_i).\label{Eq: transform}
\end{equation}
Here, $a_x=a_y=a$ and $a_z=c$ are the tetragonal lattice constants. The Hamiltonian then becomes
\begin{align}
	H_{\text{D}}(\mathbf{k}) =& c_0 + c_1\cos k_z + c_2(\cos k_x + \cos k_y) \nonumber\\
	&+ t(\sin k_x\sigma_z\tau_x - \sin k_y\tau_y) \nonumber\\
	&+ [m_0 + m_1\cos k_z + m_2(\cos k_x + \cos k_y)]\tau_z,\label{Eq: Dirac}
\end{align}
where the Pauli matrices $\sigma_i$ and $\tau_i$ represent the spin and orbital degrees of freedom, respectively. These models of DSMs preserve both time-reversal and inversion symmetries, and have a rotational symmetry axis parallel to the $z$ axis. If we confine the DSM in any direction not parallel to $z$, each of its surfaces has two disconnected Fermi arcs with opposite spin and chirality. From now on, unless stated otherwise, we choose the parameters $c_0=c_1=c_2=0$, $t = m_1 = m_2 = 1$ and $m_0=-2.5$ so that the material has a pair of Dirac points at $\mathbf{k}_D=(0,0,\pm\pi/3)$ [Fig.~\hyperref[fig:1]{1(a)}].

We will examine the WSM and DSM in a slab geometry confined in the $i$ ($i=x$ or $z$) direction, as shown in Figs.~\hyperref[fig:1]{1(b)} and \hyperref[fig:1]{1(c)}, with periodic boundary conditions (PBCs) in the other two directions. The slabs are subjected to a uniform magnetic field $\mathbf{B}\|\hat{i}$, whose magnitude can be written in terms of the magnetic flux $\Phi$ threading a unit cell as $B=\Phi/(a_ja_k)=\Phi_0\phi/(a_ja_k)$. Here, $i\neq j\neq k$, $\Phi_0=h/e$ is the magnetic flux quantum, and $\phi=\Phi/\Phi_0$.
\section{Quantum Hall effect in Weyl semimetal films}\label{sec3}
\subsection{Films confined in the $x$ direction}
\begin{figure*}
	\includegraphics[width=\linewidth]{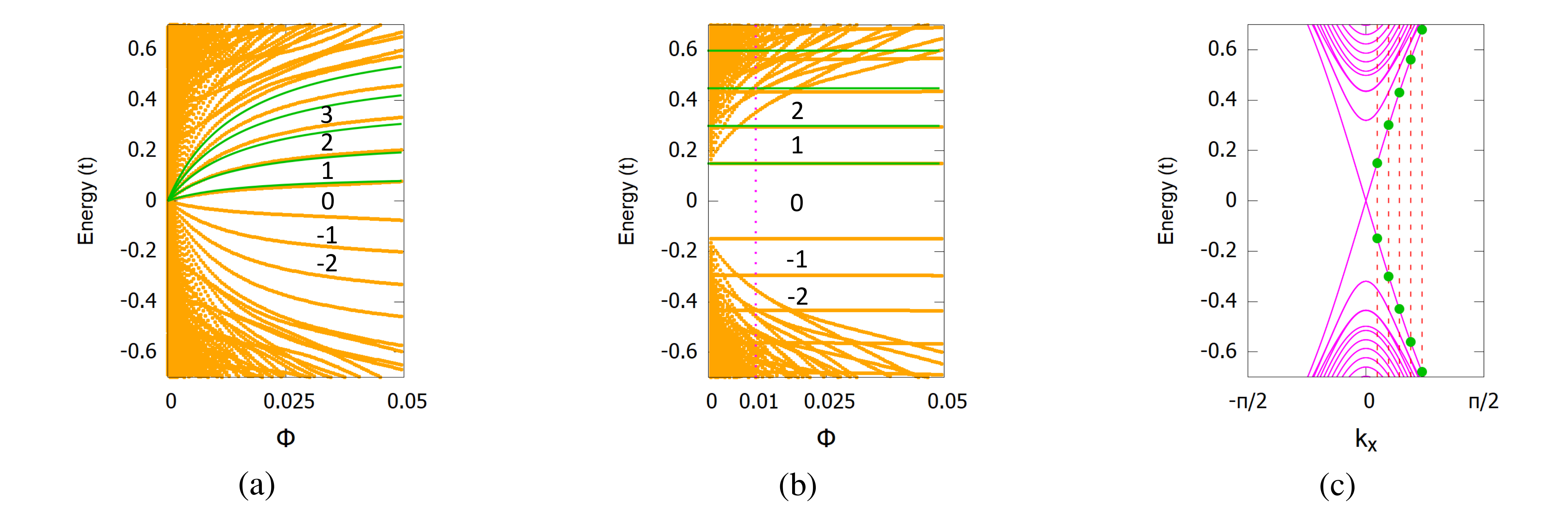}
	\caption{\label{fig:2} Energy spectra against the magnetic flux $\phi$ for (a) a WSM slab confined in the $x$ direction [Eq.~(\ref{Eq: wslabx})], and (b) a WSM bulk [Eq.~(\ref{Eq: wbulkx})] with $k_x=\pi \zeta/(N_x+1)$. The green lines represent semiclassical result determined by Eq.~(\ref{Eq: Weyl orbit}) for (a) $k_a = 2\pi/3$ and $\gamma=0.7$, (b) $k_a=0$ and $\gamma=1$. (c) The bulk Landau bands of a WSM confined in the $x$ direction and subjected to a magnetic flux $\phi=0.01$. The green dots indicate confinement subbands formed from the chiral Landau bands. The Chern numbers of some energy gaps are shown in (a) and (b). All the results are computed with $N_x=20$.}
\end{figure*}

First, we consider the WSM confined in the $x$ direction [Fig.~\hyperref[fig:1]{1(b)}] with a magnetic field $\mathbf{B}\|\hat{x}$ represented by vector potential $\mathbf{A}=(0,0,Bay)$, where $y$ is dimensionless. The Hamiltonian of our WSM is then modified in accordance with the Peierls substitution as (Appendix \ref{App:Lattice})
\begin{align}
	\mathcal{H}_{\text{W}} = \sum_{x,y,k}\frac{1}{2}\Big\{&d^{\dagger}_{xyk}[M + t'\cos(k+2\pi\phi y)]\sigma_zd_{xyk}\nonumber\\
	&+ d^{\dagger}_{xyk}(t'\sigma_z-it\sigma_x)d_{(x+1)yk} \nonumber\\
	&+ d^{\dagger}_{xyk}(t'\sigma_z-it\sigma_y)d_{x(y+1)k}\Big\} + h.c.\label{Eq: wslabx}
\end{align}
with $d^{\dagger}_{xyk}$ ($d_{xyk}$) being creation (annihilation) operator, $k$ momentum in the $z$ direction. This Hamiltonian gives an energy spectrum [Fig.~\hyperref[fig:2]{2(a)}] that was shown in Ref. \cite{Igarashi2017} with a similar model to have the energy levels in agreement with those of Weyl orbits \cite{Potter2014,Zhang2016}
\begin{equation}
	\epsilon_n = \frac{\pi t}{k_a/(2\pi\phi) + N_x+1}(n+\gamma),\quad n=0,1,\ldots\label{Eq: Weyl orbit}
\end{equation}
Here, $k_a$ is the length of the surface Fermi arcs, $N_x$ the number of lattice sites in the $x$ direction, and $\gamma$ the phase offset that depends on the distance between the Weyl points \cite{Borchmann2017}.
Additionally, to confirm the quantum Hall state of this system, we compute the Chern numbers of some energy gaps formed by the discrete levels of Weyl orbits using the Streda formula \cite{Streda1982} (Appendix \ref{App:Chern}). The nonzero Chern numbers indicate the existence of a quantized Hall conductance and are equal to the number of chiral modes at the edges of our system. Moreover, such a quantum Hall state depends on the film thickness as the energy levels do [Eq.~(\ref{Eq: Weyl orbit})], which is regarded as a signature of the quantum Hall effect based on Weyl orbits \cite{Zhang2019a,Goyal2018,Galletti2019}.

As mentioned in some experiments \cite{Zhang2017,Zhang2019a}, the Weyl orbit levels are related to the quantum confinement effect in topological semimetal films. In order to illustrate this idea, we consider those levels [Eq.~(\ref{Eq: Weyl orbit})] in the limit where the Fermi arcs vanish, $k_a=0$, and consider only the tunneling process of the quasi-particles via the bulk chiral Landau levels. In this case, we see that the energy levels take the form $\epsilon_n'=\pi t(n+1)/(N_x+1)$ and do not change with the magnetic field. On the other hand, we know that in the bulk WSM the linear dispersion of the chiral Landau levels at the Weyl points is given by
\begin{equation}
	E_c=\pm tk_x.\label{Eq: linear chiral}
\end{equation}
A comparison between $\epsilon_n'$ and $E_c$ indicates that the momentum $k_x$ is quantized in a similar way to an infinite quantum well problem, i.e., $k_x=(n+1)\pi/(N_x+1)$. This indicates that the energy levels of Weyl orbits in the limit $k_a=0$ are confinement subbands stemming from the chiral Landau levels. To further substantiate this argument, we carry out some numerical calculations as follows. We impose an additional PBC on the Hamiltonian~(\ref{Eq: wslabx}) in the $x$ direction to eliminate the surface states and preserve only the bulk spectrum. The Hamiltonian is then given by
\begin{align}
	\mathcal{H}_{\text{W}} = &\sum_{k_x,y,k_z}\frac{1}{2}\boldsymbol{\Big(}d^{\dagger}_{k_xyk_z}\big\{t\sin k_x\sigma_x + [M + t'\cos k_x \nonumber\\
	&+ t'\cos(k_z+2\pi\phi y)]\sigma_z\big\}d_{k_xyk_z} \nonumber\\
	&+ d^{\dagger}_{k_xyk_z}(t'\sigma_z-it\sigma_y)d_{k_x(y+1)k_z}\boldsymbol{\Big)} + h.c..\label{Eq: wbulkx}
\end{align}
In order to obtain only the bulk spectrum of our WSM slab, i.e., to add the effect of quantum confinement, we apply the particle-in-a-box method (PiBM) (Appendix \ref{App:confinement}) by diagonalizing this Hamiltonian only at the momenta
\begin{equation}
	k_x = \frac{\pi \zeta}{N_x+1},\qquad\zeta=1,2\ldots,N_x.\label{Eq: quantization x}
\end{equation}
The spectrum is shown in Fig.~\hyperref[fig:2]{2(b)}, where we see that distinct energy gaps still exist even in the absence of surface states. The energy levels also agree well with the semiclassical result given by Eq.~(\ref{Eq: Weyl orbit}) for $k_a=0$, as denoted by the green lines. To find their origin, we show the Landau bands obtained from Hamiltonian~(\ref{Eq: wbulkx}) for a fixed magnetic flux [Fig.~\hyperref[fig:2]{2(c)}]. Then, by adding the confinement effect [Eq.~(\ref{Eq: quantization x})], each continuous Landau band becomes a set of discrete confinement subbands. We see that the energy levels in Fig.~\hyperref[fig:2]{2(b)} are actually the confinement subbands formed from the chiral Landau levels, as we have predicted. In the presence of boundaries, these chiral Landau subbands hybridize with the surface Fermi arcs and bend towards zero energy in the low field regime, giving rise to the Weyl orbit levels shown in Fig.~\hyperref[fig:2]{2(a)}. As a result, we can conclude that the quantization of Weyl orbits gives energy levels that are chiral Landau subbands hybridizing with the surface Fermi arcs.

From this interpretation of the Weyl orbit levels, we gain two new perspectives about the QHE in topological semimetal films. First, the QHE induced by Weyl orbits is intrinsically two-dimensional instead of three-dimensional as being claimed before \cite{Nishihaya2019,Wang2017,Li2020b} since the energy levels originate from the quantum confinement effect. Further evidence for this 2D nature is that the Hall resistance of our WSM film is a factor of the Klitzing constant $R_K=h/e^2$, in agreement with the experiments, whereas in a so-claimed 3D QHE induced by the charge-density wave \cite{Tang2019a}, the Hall resistance is much smaller than $R_K$. Second, if our material somehow has the bulk chiral Landau levels but with no open Fermi arcs on its surfaces, e.g. the arcs are combined into a closed Fermi loop \cite{Potter2014}, the gaps between the chiral Landau subbands still remain. Hence, a thickness-dependent QHE or quantum oscillation in relatively thick topological semimetal films is not conclusive evidence for observing either Weyl orbits or surface Fermi arcs, in contrast to the usual expectation \cite{Zhang2019a,Goyal2018,Galletti2019,Moll2016}. Moreover, since such a QHE can take place even without the surface Fermi arcs, a question then naturally arises; is the QHE observable in our WSM when it is confined in the $z$ direction, which has no nontrivial states on its boundaries?
\subsection{Films confined in the $z$ direction}\label{sec: 3B}
\begin{figure}
	\includegraphics[width=\linewidth]{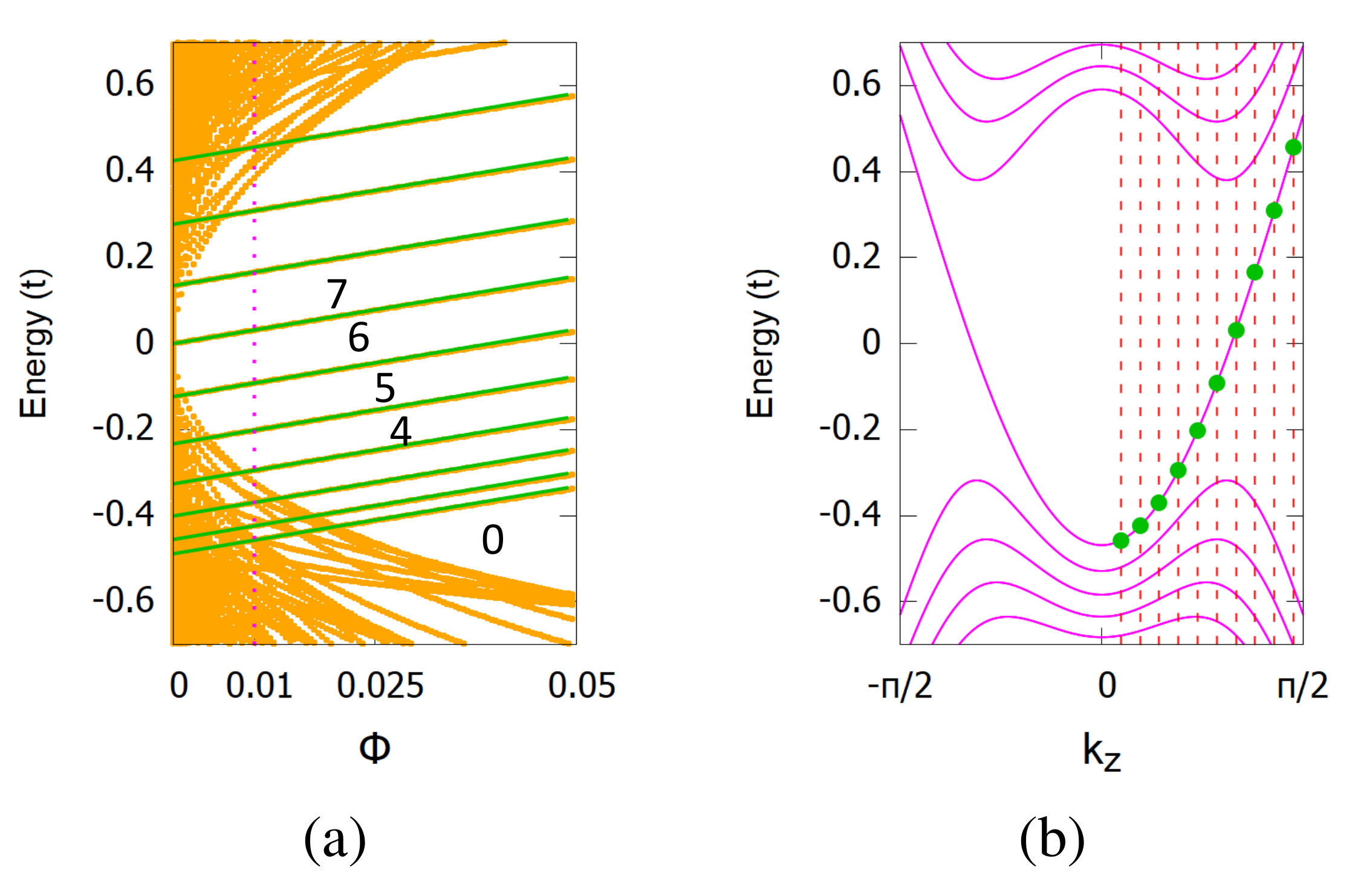}
	\caption{\label{fig:3} (a) Energy spectrum against the magnetic flux $\phi$ for a WSM slab confined in the $z$ direction. The green lines represent energy levels determined by Eq.~(\ref{Eq: level z}). The Chern numbers of some energy gaps are shown. (b) The bulk Landau bands of a WSM confined in the $z$ axis and subjected to a magnetic flux $\phi=0.01$. The green dots indicate confinement subbands formed from the chiral Landau band. All the results are computed with $N_z=20$.}
\end{figure}

We now consider a WSM confined in the $z$ direction [Fig.~\hyperref[fig:1]{1(c)}] with a magnetic field $\mathbf{B}\|\hat{z}$ given by vector potential $\mathbf{A}=(0,Bax,0)$. Similar to the previous case, the bulk spectrum of our WSM decomposes into 1D Landau bands dispersing along the $z$ direction, as shown in Fig.~\hyperref[fig:3]{3(b)}, and the chiral level still evolves differently from other bands. If we introduce confinement in $z$, we expect that the chiral Landau subbands also form energy gaps distinct from others, and the QHE in relatively thick films will thus be observable.
The Hamiltonian of our system now reads
\begin{align}
	\mathcal{H}_{\text{W}} =&\sum_{x,k,z}\frac{1}{2}\boldsymbol{\Big(}d^{\dagger}_{xkz}\big\{[M + t'\cos(k+2\pi\phi x)]\sigma_z \nonumber\\
	&+ t\sin(k+2\pi\phi x)\sigma_y\big\}d_{xkz} + d^{\dagger}_{xkz}t'\sigma_zd_{xk(z+1)} \nonumber\\
	&+ d^{\dagger}_{xkz}(t'\sigma_z-it\sigma_x)d_{(x+1)kz}\boldsymbol{\Big)} + h.c.\label{Eq: HWz}
\end{align}
with $k$ being momentum in the $y$ direction. The energy spectrum of this Hamiltonian is shown in Fig.~\hyperref[fig:3]{3(a)}. Since this slab does not have topological surface states, we can also obtain its spectrum analytically by using the PiBM. We start with adding the effect of magnetic field $\mathbf{B}$ on the continuum model [Eq.~(\ref{Eq: Weylconti})], which is done by replacing the momenta $k_x$ and $k_y$ with ladder operators $l$, $l^{\dagger}$ as
\begin{equation}
	k_x\rightarrow \sqrt{\pi\phi}(l^{\dagger}+l),\quad k_y\rightarrow -i\sqrt{\pi\phi}(l^{\dagger}-l).\label{eq: ladder}
\end{equation}
The Hamiltonian then becomes
\begin{equation}
	H_{\text{W}}(k_z) = \begin{pmatrix}
		\mathcal{M}(k_z) & 2\sqrt{\pi\phi}tl\\ 2\sqrt{\pi\phi}tl^{\dagger} & -\mathcal{M}(k_z)
	\end{pmatrix},
\end{equation}
with $\mathcal{M}(k_z) = m - t'k_z^2/2 - 2\pi\phi t'\left(l^{\dagger}l + \dfrac{1}{2}\right)$. Then, using the trial wavefunctions $(\alpha_1\ket{\nu-1}, \alpha_2\ket{\nu})^T$ for $\nu=1,2,\ldots$ and $(0,\ket{0})^T$ for $\nu=0$, where $\nu$ is the band index, we can obtain the spectrum of $H_{\text{W}}(k_z)$ from the secular equations as
\begin{equation}
	\det\begin{vmatrix}
		K_{\nu k_z} + \pi t'\phi - E & 2\sqrt{\pi\phi}t\sqrt{\nu}\\
		2\sqrt{\pi\phi}t\sqrt{\nu} & -K_{\nu k_z} + \pi t'\phi - E
	\end{vmatrix} = 0,
\end{equation}
for $\nu=1,2,\ldots$, and
\begin{equation}
	-K_{(\nu=0)k_z} + \pi t'\phi - E = 0,\qquad\text{for }\nu=0.
\end{equation}
Here, $K_{\nu k_z} = m - \frac{t'}{2}k_z^2 - 2\pi t'\phi \nu$. The 1D Landau bands of the WSM are then given by
\begin{align}
	E_0(k_z) &= -m + \frac{t'}{2}k_z^2 + \pi t'\phi,\\
	E_{\nu}(k_z) &= \pm\sqrt{K^2_{\nu k_z} + 4\pi t^2\phi \nu} + \pi t'\phi.
\end{align}
Now, we can transform the dispersion of the zeroth level into a lattice version by substituting $k_z^2\rightarrow2(1-\cos k_z)$, which yields
\begin{equation}
	E_0(k_z) = -M - 2t' - t'\cos k_z + \pi t'\phi.\label{Eq: quadratic chiral}
\end{equation}
Finally, we take into account the effect of quantum confinement by employing the PiBM, i.e., replacing $k_z=\pi\zeta/(N_z+1)$ with $\zeta=1,2,\ldots,N_z$. The subbands of chiral Landau level read
\begin{equation}
	\epsilon_0(\zeta) = -M - 2t' - t'\cos\frac{\pi\zeta}{N_z+1} + \pi t'\phi,\label{Eq: level z}
\end{equation}
and they evolve linearly with respect to the magnetic flux $\phi$, as shown by the green lines in Fig.~\hyperref[fig:3]{3(a)}. On the other hand, the Landau bands with $\nu>0$ move away from $0$ and make the gaps between the chiral Landau subbands observable. These gaps also have nonzero Chern numbers, indicating the existence of the QHE.
\section{Discussion}\label{sec4}
Based on the spectrum in Fig.~\hyperref[fig:3]{3(a)}, we make two inferences:

First, we see that when our WSM is confined in the $z$ direction, the dependence of its chiral Landau subbands on $\phi$ deviates considerably from that of the slab perpendicular to the $x$ axis. In particular, if confinement is in the $x$ direction, and we neglect the surface Fermi arcs, the subbands stay constant as the field strength increases [Fig.~\hyperref[fig:2]{2(b)}], which agrees with the semiclassical equation. On the other hand, if the WSM is confined in $z$, the subbands depend linearly on the flux and are unevenly spaced, and thus behave differently from Eq.~(\ref{Eq: Weyl orbit}). From Eq.~(\ref{Eq: level z}), we know that such a difference stems from the second-order terms in $k$ of Hamiltonian~(\ref{Eq: Weylconti}) while Eq.~(\ref{Eq: Weyl orbit}) was obtained by using a linear dispersion of the chiral Landau levels \cite{Potter2014}. However, another problem then comes up: those quadratic terms in $k$ contribute substantially to the spectrum in Fig.~\hyperref[fig:3]{3(a)} but does not affect the one in Fig.~\hyperref[fig:2]{2(b)}. We investigate this problem and revisit the semiclassical Weyl orbits in subsection \ref{subsec: 1}.

Second, since a DSM can be regarded as a combination of two WSMs with opposite spin and chirality, we expect that a $(001)$ film of DSM will have a spectrum with nonzero spin Hall conductance. We demonstrate this idea in subsection \ref{subsec: 2}, and show that such a quantum spin Hall effect (QSHE) is observable in the DSM candidate Cd$_3$As$_2$.
\subsection{Dependence of the chiral Landau levels on the $k^2$-terms in $H_{\text{W}}(\mathbf{k})$}\label{subsec: 1}
We study how the $k^2$-terms in Hamiltonian~(\ref{Eq: Weylconti}) affect the dispersion of the zeroth Landau bands by finding their analytical expressions when the magnetic field $\mathbf{B}$ is applied along an arbitrary direction. We transform the vectors $(k_x,k_y,k_z)$ of the crystal frame into the $(k_1,k_2,k_3)$ of the magnetic frame ($\hat{k}_3\|\mathbf{B}$) using a 3D rotation matrix. For simplicity and without loss of generality, we assume that $\mathbf{B}\perp\hat{k}_y$ and rotate the vectors about the $y$ axis ($k_2\equiv k_y$) as
\begin{equation}
	\begin{pmatrix} k_1 \\ k_2 \\ k_3 \end{pmatrix} = 
	\begin{pmatrix}
		\cos\theta & 0 & -\sin\theta\\
		0 & 1 & 0\\
		\sin\theta & 0 & \cos\theta
	\end{pmatrix}\begin{pmatrix}
		k_x \\ k_y \\ k_z
	\end{pmatrix}
\end{equation}
with $\theta$ being the angle between $k_3$ and $k_z$. The continuum Hamiltonian of our WSM then reads
\begin{equation}
	H_{\text{W}}(\mathbf{k}) = t(\cos\theta k_1 + \sin\theta k_3)\sigma_x + k_2\sigma_y + \mathcal{M}(\mathbf{k})\sigma_z ,\label{Eq: H rotate}
\end{equation}
where $\mathcal{M}(\mathbf{k}) = m - t'(k_1^2+k_2^2+k_3^2)/2$. The eigenvalues of this Hamiltonian are given by $E = \pm\sqrt{\mathcal{M}^2 + t^2[(\cos\theta k_1+\sin\theta k_3)^2+k_2^2]}$, and the Weyl points are located at $(-k_W\sin\theta,0,k_W\cos\theta)$ and $(k_W\sin\theta,0,-k_W\cos\theta)$, $k_W=\sqrt{2m/t'}$.
\subsubsection{Bulk chiral Landau levels}
To find the chiral Landau bands analytically, we consider the parameters $\{t,t',m\}$ that satisfy the constraint $t'k_W/t=\eta=\pm1$, which keeps the velocity at the Weyl points isotropic. In the vicinity of the first Weyl point $(-k_W\sin\theta,0,k_W\cos\theta)$, the Hamiltonian reads
\begin{align}
	&H_{\text{W}}(\mathbf{k}') = tk_1'(\cos\theta\sigma_x + \eta\sin\theta\sigma_z) + tk_2\sigma_y \nonumber\\
	&+ tk_3'(\sin\theta\sigma_x - \eta\cos\theta\sigma_z) - \frac{t'}{2}(k_1'^2+k_2^2+k_3'^2)\sigma_z,
\end{align}
where $k_1'=k_1+k_W\sin\theta$ and $k_3'=k_3-k_W\cos\theta$. When a magnetic field $\mathbf{B}$ is applied, the momenta $k_1$ and $k_2$ are quantized in terms of the ladder operators similar to Eq.~(\ref{eq: ladder}). We then have $H_{\text{W}}(k_z) = H_1(k_z) + H_2(k_z)$, where the first-order term is
\begin{align}
	&H_1(k_z) = tk_3'\begin{pmatrix} -\eta\cos\theta & \sin\theta \\ \sin\theta & \eta\cos\theta\end{pmatrix} \nonumber\\
	&+ t\sqrt{\pi\phi}\begin{pmatrix}
		\eta\sin\theta(l^{\dagger}+l) & \cos\theta(l^{\dagger}+l) - (l^{\dagger}-l) \\
		\cos\theta(l^{\dagger}+l) + (l^{\dagger}-l) & -\eta\sin\theta(l^{\dagger}+l)
	\end{pmatrix},
\end{align}
and the second-order one reads
\begin{equation}
	H_2(k_z) = -\frac{t'}{2}\left[2\pi\phi(2l^{\dagger}l + 1) + k_3'^2\right]\sigma_z.
\end{equation}
We now make an approximation by solving the chiral Landau level from $H_1$ and treating $H_2$ as a perturbation. With the eigenvector $\ket{L_{c+}} = [1/\sqrt{2(1+\cos\theta)}](\eta\sin\theta, 1+\cos\theta)^T\ket{0}$, the chiral Landau level at the first Weyl point is given by $E_+^{(0)}(k_3') = \eta tk_3'$. The first-order correction from the perturbation $H_2$ is
\begin{equation}
	E_+^{(1)}(k_3') = \braket{L_{c1}|H_2|L_{c1}} = t'\left(\pi\phi + \frac{k_3'^2}{2}\right)\cos\theta .
\end{equation}
The dispersion of the zeroth Landau level is then expressed as
\begin{eqnarray}
	E_+(k_3) &=& \eta t(k_3-k_W\cos\theta) \nonumber\\
	&&+ t'\left[\pi\phi + \frac{(k_3-k_W\cos\theta)^2}{2}\right]\cos\theta .\label{Eq: lls +}
\end{eqnarray}
The first-order correction from $H_2$ allows us to reproduce both Eq.~(\ref{Eq: linear chiral}) and Eq.~(\ref{Eq: quadratic chiral}), and hence we neglect all higher-order corrections.

\begin{figure*}
	\includegraphics[width=\linewidth]{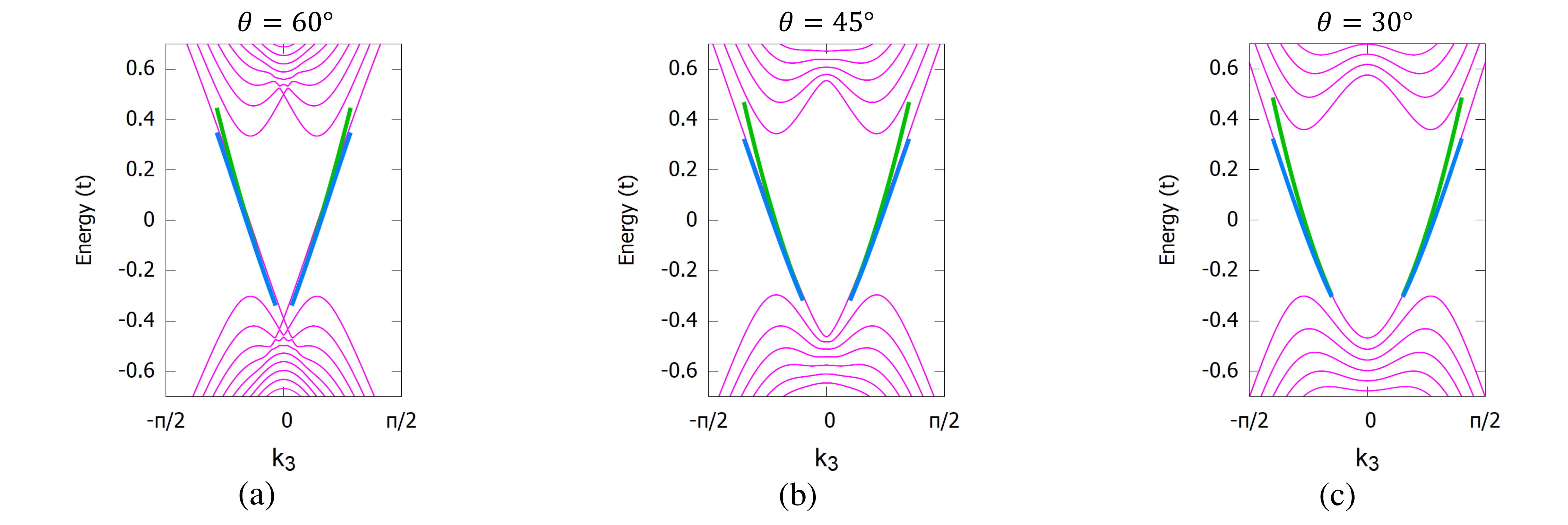}
	\caption{\label{fig:4} Landau bands of our WSM bulk in a magnetic field with (a) $\theta=60^{\circ}$, (b) $\theta=45^{\circ}$, and (c) $\theta=30^{\circ}$. The Landau bands (magenta lines) are obtained from the lattice model of Hamiltonian (\ref{Eq: H rotate}). The green lines represent the analytical results given by Eqs.~(\ref{Eq: lls +}) and (\ref{Eq: lls -}). The blue lines are also those equations but being modified to a lattice version by Eq.~(\ref{Eq: transform}). The green and blue lines approach each other as the Weyl points come close to the $\Gamma$-point.}
\end{figure*}

A similar calculation yields the chiral Landau level crossing the second Weyl point $(-\sin k_W,0,\cos k_W)$ as
\begin{eqnarray}
	E_-(k_3) &=& -\eta t(k_3+k_W\cos\theta) \nonumber\\
	&&+ t'\left[\pi\phi + \frac{(k_3+k_W\cos\theta)^2}{2}\right]\cos\theta .\label{Eq: lls -}
\end{eqnarray}
These expressions demonstrate that the contribution from $H_2$ to the chiral Landau levels vanishes when the magnetic field is perpendicular to the line connecting the two Weyl points, i.e., $\mathbf{B}\|\hat{x}$. Besides, the effect of $H_2$ becomes more significant when the magnetic field direction approaches the $z$ axis. To assess the reliability of these results, we compare Eqs.~(\ref{Eq: lls +}) and (\ref{Eq: lls -}) with the Landau bands obtained from the lattice model of Eq.~(\ref{Eq: H rotate}), as shown in Fig.~\hyperref[fig:4]{4}.
\subsubsection{QHE in a WSM slab confined in an arbitrary direction}
After getting a more general expression for the chiral Landau bands to explain the difference between Figs.~\hyperref[fig:2]{2(b)} and \hyperref[fig:3]{3(a)}, it is interesting to seek an expression for the Weyl orbit levels taking into account the effect of the $k^2$-terms. The Onsager - Bohr - Sommerfeld quantization for a classical orbit reads
\begin{equation}
	\oint\mathbf{p}\cdot d\mathbf{r} = 2\pi\hbar(n+\gamma).
\end{equation}
After some calculations, we get
\begin{align}
	\epsilon_n = \frac{B}{t'\cos\theta}\left[A + B - \sqrt{2AB + B^2 + t^2 - 2\pi\phi t'^2\cos^2\theta}\right]\label{Eq: minh Weyl orbits}
\end{align}
with
\begin{equation}
	A = t'\left[\frac{\pi(n+\gamma)}{L_z'}\cos\theta + k_W\sin^2\theta\right]\text{ and } B = 2\pi\phi t\frac{L_z'}{k_a}.
\end{equation}
Here, $L_z'=(N_z+1)/\cos(\Theta-\theta)$ is dimensionless, $k_a = 2k_W\sin\Theta$, $\theta$ and $\Theta$ determine the directions of magnetic field and confinement, respectively. In the limits $\Theta=\theta\rightarrow 90^{\circ}$ and $\Theta=\theta\rightarrow 0^{\circ}$, this equation reproduces Eq.~(\ref{Eq: Weyl orbit}) and Eq.~(\ref{Eq: level z}).
\begin{figure*}
	\includegraphics[width=\linewidth]{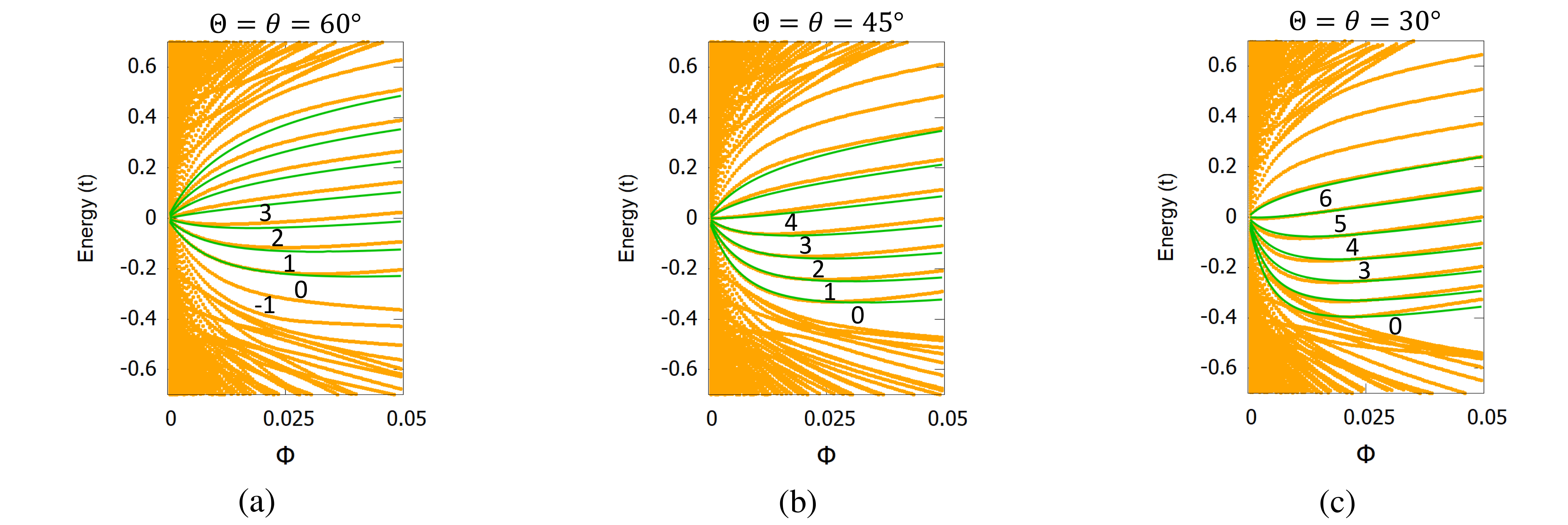}
	\caption{\label{fig:5} Energy spectra against the magnetic flux $\phi$ for a WSM slab confined in the $k_3$ direction determined by (a) $\Theta=60^{\circ}$, (b) $\Theta=45^{\circ}$, and (c) $\Theta=30^{\circ}$. The green lines represent semiclassical result determined by Eq.~(\ref{Eq: minh Weyl orbits}) with $\gamma=0.7$. The Chern numbers of some energy gaps are shown. All the results are computed with a thickness corresponding to $N_3=20$.}
\end{figure*}

We now compare Eq.~(\ref{Eq: minh Weyl orbits}) with the results obtained from the lattice model of Hamiltonian (\ref{Eq: H rotate}). A WSM confined in the $k_3$ direction and subjected to a perpendicular magnetic field is described by
\begin{align}
	\mathcal{H}_{\text{W}} = &\sum_{x_1,k,x_3}\frac{1}{2}\boldsymbol{\Big(}a^{\dagger}_{x_1kx_3}\Big\{ t\sin (k+2\pi\phi x_1)\sigma_y \nonumber\\
	&+ \big[M + t'\cos(k+2\pi\phi x_1)\big]\sigma_z\Big\} a_{x_1kx_3}\nonumber\\
	&+ a^{\dagger}_{x_1kx_3}\left(t'\sigma_z - it\cos\theta\sigma_x\right)a_{(x_1+1)kx_3}\nonumber\\
	&+ a^{\dagger}_{x_1kx_3}\left(t'\sigma_z - it\sin\theta\sigma_x\right)a_{x_1k(x_3+1)}\boldsymbol{\Big)} + h.c. .
\end{align}
The energy spectra of this WSM are shown in Fig.~\hyperref[fig:5]{5} for different tilting angles, which shows a transition from Fig.~\hyperref[fig:2]{2(a)} to Fig.~\hyperref[fig:3]{3(a)}, and agree well with Eq.~(\ref{Eq: minh Weyl orbits}). When the confinement direction deviates from the $x$ axis, both the spacing and the dependence on magnetic flux $\phi$ of the chiral Landau subbands change. Interestingly, we can see how the subbands evolve by computing the Chern numbers of the gaps between them.

In DSMs, we assume that the physics is somewhat similar. As the experiments about Weyl orbits are often conducted in $(112)$ films of Cd$_3$As$_2$, this result may give better explanations to those magneto-transport studies, e.g. the high values of the Landau indices in the quantum Hall measurements.
\subsection{Quantum spin Hall effect in topological Dirac semimetal thin films}\label{subsec: 2}
\begin{figure}
	\includegraphics[width=\linewidth]{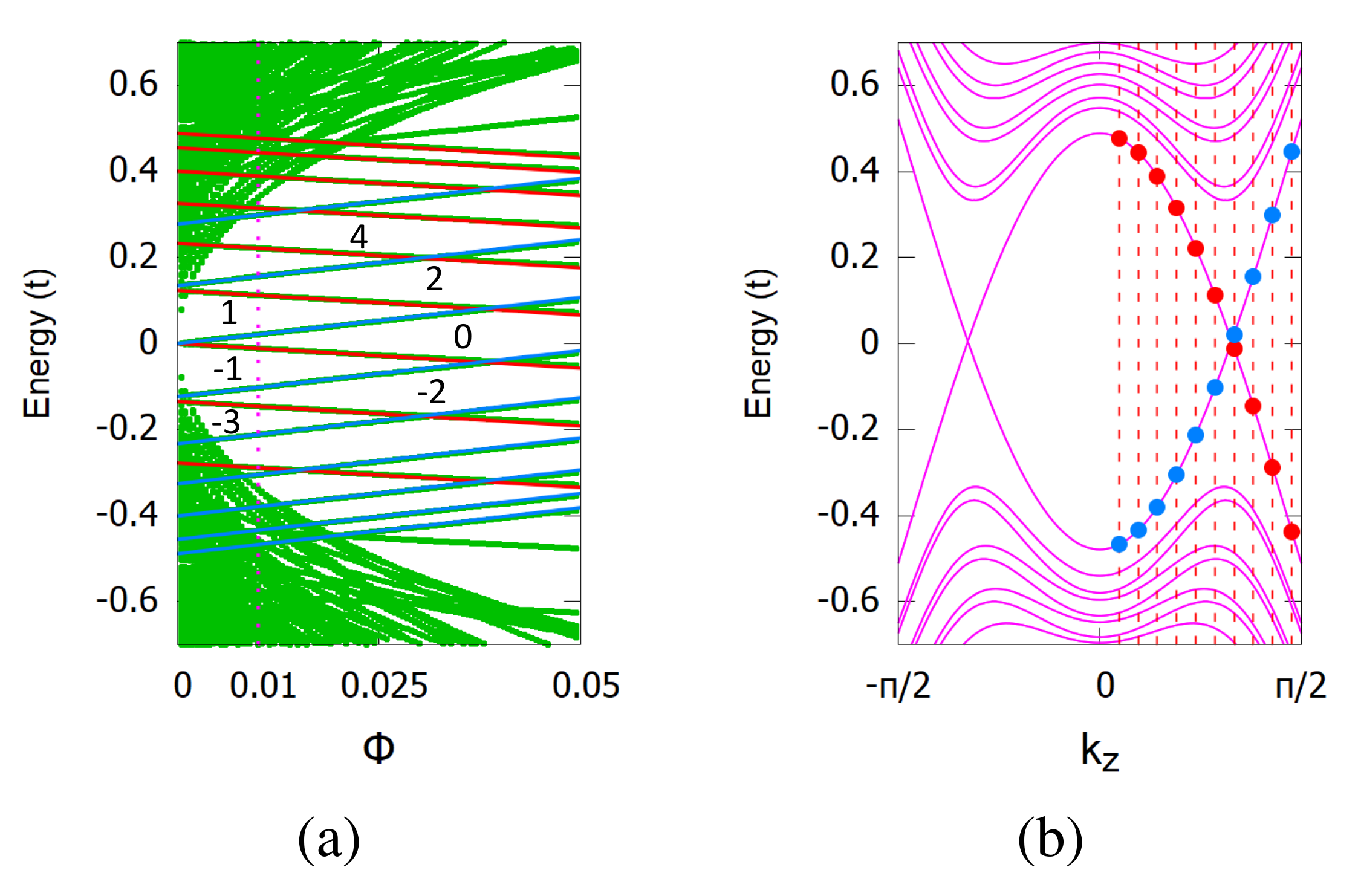}
	\caption{\label{fig:6} (a) Energy spectrum against the magnetic flux $\phi$ for a DSM slab. The red lines denote spin-up chiral Landau subbands given by Eq.~(\ref{Eq: Dirac subbands up}). The blue lines denote spin-down chiral Landau subbands given by Eq.~(\ref{Eq: Dirac subbands down}). The Chern numbers of some energy gaps are shown. (b) Bulk Landau bands of the DSM confined in the $z$ axis and subjected to a magnetic flux $\phi=0.01$. The red and blue dots indicate confinement subbands formed from the chiral Landau bands. All the results are computed with $N_z=20$.}
\end{figure}

Finally, we show how the QHE and QSHE induced by chiral Landau subbands take place in the DSM films. For our DSM [Eq.~(\ref{Eq: Dirac})], a magnetic field $\mathbf{B}\|\hat{z}$ preserves the spin-$z$ component ($S_z$) and lifts the spin degeneracy even in the absence of Zeeman coupling. A DSM film grown along its rotational symmetry axis ($z$) and subjected to $\mathbf{B}$ is described by
\begin{align}
	\mathcal{H}_{\text{D}} = \sum_{x,k,z}\frac{1}{2}\Big\{&d^{\dagger}_{xkz}\big[m_0\tau_z + m_2\cos(k+2\pi\phi x)\tau_z \nonumber\\
	&- t\sin(k+2\pi\phi x)\tau_y + \lambda_z\phi\sigma_z\tau_g\big]d_{xkz} \nonumber\\
	&+ d^{\dagger}_{xkz}(m_2\tau_z - it\sigma_z\tau_x)d_{(x+1)kz}\nonumber\\
	& + d^{\dagger}_{xkz}m_1\tau_zd_{xk(z+1)}\Big\} + h.c.,
\end{align}
where $\lambda_z=\dfrac{h^2}{8\pi m_ea^2}$ and
\begin{equation}
	\tau_g = \begin{pmatrix}
		g_s & 0 \\ 0 & g_p
	\end{pmatrix},
\end{equation}
$g_s$ and $g_p$ are the effective $g$-factors for $s$ and $p$ orbitals, respectively. For simplicity, we choose $\lambda_zg_s=2$ and $\lambda_zg_p=1$. Solving the eigenvalues of $\mathcal{H}_D$ numerically gives an energy spectrum as shown in Fig.~\hyperref[fig:6]{6(a)}, which is composed of two sets of chiral Landau subbands with opposite chirality and spin polarization. They evolve in different directions with respect to the magnetic field and thus cross each other to form the energy gaps that can be seen as an overlap of two separate gaps. This spectrum is also explained by the quantum confinement picture as shown in Fig.~\hyperref[fig:6]{6(b)}. According to this figure, we should notice that the two sets of subbands cross if $\pi/(N_z+1)<|\mathbf{k}_D|$, i.e.,
\begin{equation}
	L_z>\frac{\pi}{|\mathbf{q}_D|}.\label{Eq: thickness}
\end{equation}
To find how the chiral Landau subbands depend on the magnetic field, we also follow the same calculation as presented in Sec.~\ref{sec: 3B} and get an analytical expression of those subbands
\begin{subequations}
	\begin{eqnarray}
		\epsilon_{0\uparrow}(q_z) = C_0 &+& M_0 + (C_1+M_1)q_z^2 \nonumber\\
		&+& (C_2+M_2)\frac{eB}{\hbar} + \frac{\mu_Bg_sB}{2},
	\end{eqnarray}
	\begin{eqnarray}
		\epsilon_{0\downarrow}(q_z) = C_0 &-& M_0 + (C_1-M_1)q_z^2 \nonumber\\
		&+& (C_2-M_2)\frac{eB}{\hbar} - \frac{\mu_Bg_pB}{2}.
	\end{eqnarray}
\end{subequations}
To make a comparison with the lattice model spectrum [Fig.~\hyperref[fig:6]{6(a)}], we let $C_i=0$ and transform the equations using Eq.~(\ref{Eq: transform}), which gives
\begin{subequations}
	\label{eq:whole}
	\begin{equation}
		\epsilon_{0\uparrow}(\zeta) = m_0 + 2m_2 + m_1\cos\frac{\pi\zeta}{N_z+1} - \pi m_2\phi + \lambda_zg_s\phi,\label{Eq: Dirac subbands up}
	\end{equation}
	\begin{equation}
		\epsilon_{0\downarrow}(\zeta) = -m_0 - 2m_2 - m_1\cos\frac{\pi\zeta}{N_z+1} + \pi m_2\phi - \lambda_zg_p\phi.\label{Eq: Dirac subbands down}
	\end{equation}
\end{subequations}
These expressions show that the presence of Zeeman interaction still keeps the chiral Landau subbands evolving linearly with the magnetic flux but changes their slopes. Specifically, the $g_s$-factor modifies the slope of the spin-up chiral Landau subbands whereas $g_p$ affects the spin-down ones, which reflects the band inversion of DSMs.

Due to  our choice of parameters, the spin-up chiral Landau subbands are hole-like, and the spin-down ones are electron-like. As a result, when our material has an additional boundary, the spin-up subbands are bent downward and give right-handed edge states while those with spin-down disperse upward giving left-handed edge states. The combination of these edge modes may give rise to the coexistence of chiral and helical edge states. For example, we consider an energy gap of Chern number $\mathcal{C}=1$, which is a combination of $\mathcal{C}_{\downarrow}=7$ and $\mathcal{C}_{\uparrow}=-6$ gaps. When the Fermi level lies within this gap, the spin Hall conductance is quantized as \cite{Murakami2006,Yang2006}
\begin{equation}
	\sigma_s = -C_s\frac{e}{2\pi}
\end{equation}
with $C_s=(C_{\uparrow}-C_{\downarrow})/2=-6.5$. At the boundary, energy levels are bent and form $7$ left-handed spin-down and $6$ right-handed spin-up edge states, which would result in a total of $1$ chiral edge state and $6$ helical edge states. Hence, both quantum Hall and quantum spin Hall phases coexist in our system whose time-reversal symmetry is broken by the magnetic field. In this case, the helical edge states exist because $S_z$ is approximately conserved \cite{Zhang2014}. This scenario is well-known in graphene \cite{Abanin2006,Fertig2006,Young2014}, where the helical edge states are protected by an additional symmetry instead of the time-reversal symmetry as in topological insulators.
\subsubsection*{Estimation for Cd$_3$As$_2$}
\begin{figure}
	\includegraphics[width=\linewidth]{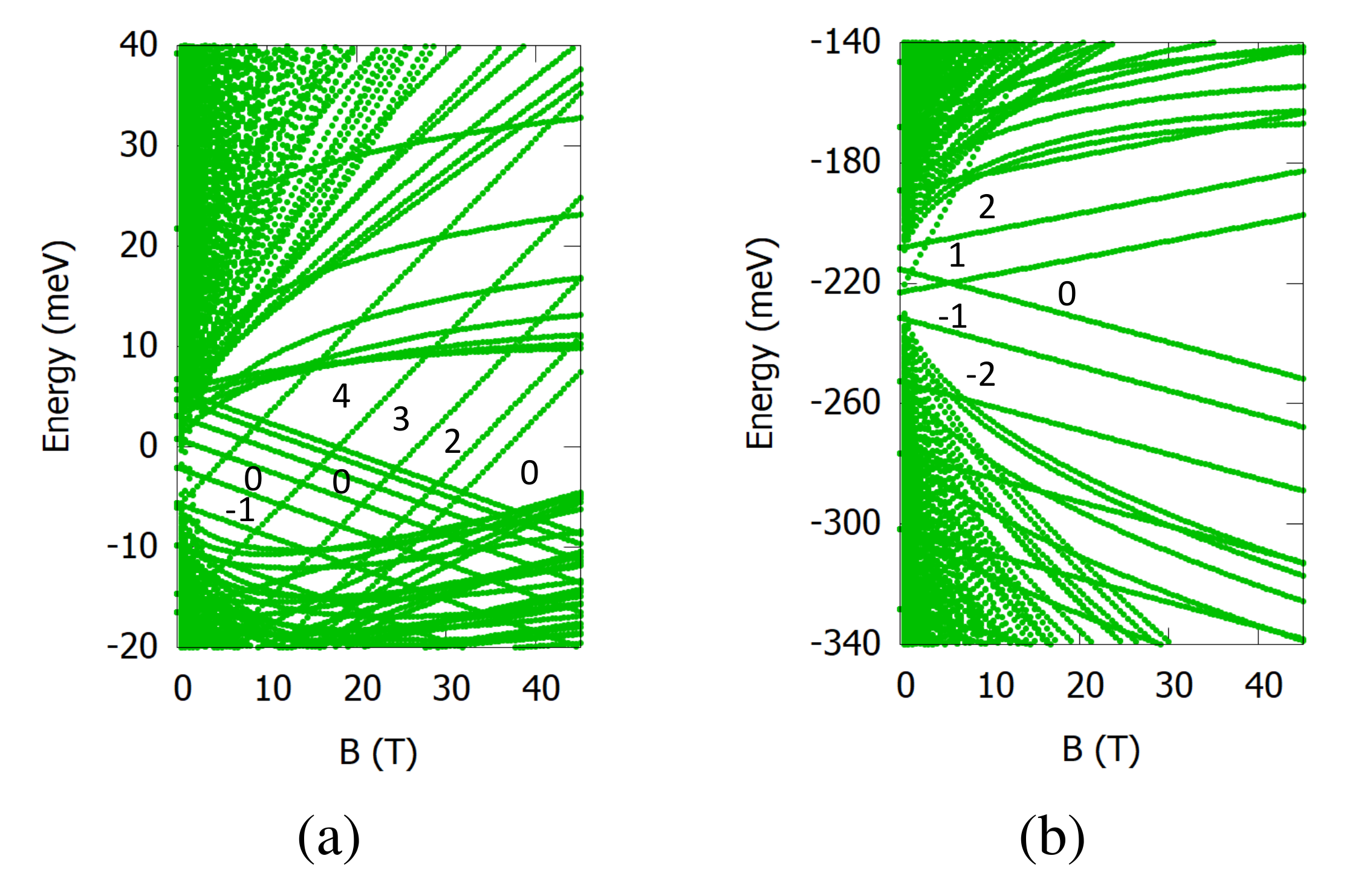}
	\caption{\label{fig:7}Energy spectra against the magnetic flux $\phi$ for a $(001)$ film of Cd$_3$As$_2$ described by two different models: (a) Hamiltonian (\ref{Eq: Diracconti}) with parameters given by Ref. \cite{Cano2017}, the film thickness is $50$ nm, and (b) the hyperbolic model proposed by Ref. \cite{Jeon2014}, the film thickness is $120$ nm. Some Chern numbers are shown in both figures by comparing with Fig.~\hyperref[fig:6]{6(a)}.}
\end{figure}

Since a quantized Hall conductance has recently been observed in $(001)$ films of Cd$_3$As$_2$ \cite{Kealhofer2020} under a strong magnetic field, we roughly estimate whether the QHE and QSHE induced by the chiral Landau subbands can be observed in such films. Depending on the growth condition, Cd$_3$As$_2$ can be either centrosymmetric or noncentrosymmetric \cite{Sankar2015b}. Here, we consider a centrosymmetric Cd$_3$As$_2$, which is more popular and can be described by Hamiltonian~(\ref{Eq: Diracconti}) with parameters obtained by fitting with the \textit{ab initio} calculation \cite{Cano2017}. Additionally, as the distance between the Dirac points of Cd$_3$As$_2$ obtained by the Landau level spectroscopy measurements \cite{Akrap2016,Jeon2014,Desrat2018,Krizman2019} is one order smaller than the \textit{ab initio} calculations \cite{Wang2013,Ali2014a,Conte2017a}, we also consider the hyperbolic model proposed by Ref. \cite{Jeon2014}. A detailed procedure is presented in Appendix \ref{App:quantization}. The spectra computed for these two models are shown in Fig.~\hyperref[fig:7]{7}, where, in both cases, we use $g_s=18.6$ \cite{Jeon2014} and choose $g_p=10$ for simplicity.

In general, both spectra have three types of gaps formed by the chiral Landau subbands: those with nonzero spin Chern number, those with nonzero Chern number and zero spin Chern number, and one with vanishing Chern and spin Chern numbers. We are interested in the gaps with nonzero spin Chern number, which do not exist if the film is too thin but is not observable if the film is too thick. If we assume an energy gap $\Delta\epsilon\ge10$~K $\sim 1$ meV to be observable, the majority of gaps in Fig.~\hyperref[fig:7]{7(a)} and all gaps in Fig.~\hyperref[fig:7]{7(b)} satisfy the condition. Hence, the QHE induced by the chiral Landau subbands can be observed in relatively thick films compared to the conventional quantum wells ($5$-$20$ nm).

If we use this simple model [Fig.~\hyperref[fig:7]{7(a)}] to explain the QHE observed in Ref. \cite{Kealhofer2020}, it may capture two features of the experiment, i.e., the absence of $\nu=1$ plateau and the high resistance at strong magnetic field. Nevertheless, just like the explanation given in that reference, it is also unable to clarify the difference in activation energy between the states at even and odd filling factors. On the other hand, if the actual distance between the Dirac nodes agrees with the Landau level spectroscopy measurements, the QHE induced by chiral Landau subbands may not be observable in such a quantum Hall experiment since the film thickness does not satisfy Eq.~(\ref{Eq: thickness}).
\section{Conclusion}\label{sec5}
Our study gives a generic and simplified picture of the QHE induced by confinement subbands stemming from the chiral Landau levels in topological semimetal films. Using a minimal model of WSM, we have demonstrated that the energy levels of Weyl orbits originate from the confinement subbands of the chiral Landau levels that hybridize with the surface Fermi arcs. We have then studied how the $k^2$-terms in a WSM Hamiltonian affect the evolution of the chiral Landau bands with respect to the magnetic field. We have found a general expression for the Weyl orbit levels and shown how the QHE takes place in a WSM film confined in an arbitrary direction. Furthermore, when examining a DSM confined in its rotational symmetry axis, we may have not only explained the QHE recently observed but also predicted the coexistence of both quantum Hall and quantum spin Hall states in such a system.

\begin{acknowledgements}
	This work was supported by the Japan Society for the Promotion of Science KAKENHI (Grant Nos.
	JP19K14607 and JP20H01830) and by CREST, Japan Science and Technology Agency (Grant No. JPMJCR18T2).
\end{acknowledgements}

\appendix
\section{\label{App:Lattice}Lattice Hamiltonian in a magnetic field}
We present how to obtain the Hamiltonian in Eq.~(\ref{Eq: wslabx}) and write its explicit form. First, we take the Fourier transform of the lattice model (\ref{Eq: Weyl}) into real space as
\begin{align}
	&\mathcal{H}_{\text{W}} = -\frac{1}{2}\sum_{\mathbf{m}}\Big[d^{\dagger}_{\mathbf{m}}(t'\sigma_z-it\sigma_x)d_{\mathbf{m}+\hat{x}} + d^{\dagger}_{\mathbf{m}+\hat{x}}(t'\sigma_z +\nonumber\\
	&it\sigma_x)d_{\mathbf{m}} + d^{\dagger}_{\mathbf{m}}(t'\sigma_z-it\sigma_y)d_{\mathbf{m}+\hat{x}} + d^{\dagger}_{\mathbf{m}+\mathbf{1}_y}(t'\sigma_z+it\sigma_y)d_{\mathbf{m}}\nonumber\\
	& - 4d^{\dagger}_{\mathbf{m}}M\sigma_zd_{\mathbf{m}} + d^{\dagger}_{\mathbf{m}}t'\sigma_zd_{\mathbf{m}+\hat{z}} + d^{\dagger}_{\mathbf{m}+\hat{z}}t'\sigma_zd_{\mathbf{m}}\Big],\label{weyl:real}
\end{align}
where $\mathbf{m}=(x,y,z)$ is a dimensionless position vector. In the presence of a magnetic field $\mathbf{B}$, the Hamiltonian is modified by the Peierls substitution so that it remains invariant under the $U(1)$ gauge transformation. In particular, the hopping integrals are changed, such as
\begin{equation}
	d^{\dagger}_{\mathbf{m}+\hat{y}}td_{\mathbf{m}} \longrightarrow d^{\dagger}_{\mathbf{m}+\hat{y}}t\exp(i\vartheta^{(y)}_{\mathbf{m}})d_{\mathbf{m}}
\end{equation}
with $\vartheta^{(y)}_{\mathbf{m}} = -\frac{2\pi}{\Phi_0}\int_{\mathbf{m}}^{\mathbf{m}+\hat{y}}\mathbf{A}\cdot\hat{y}ady$. Using the vector potential $\mathbf{A}=(0,0,Bay)$, we get $\vartheta^{(x)}_{\mathbf{m}} = \vartheta^{(y)}_{\mathbf{m}} = 0$ and $\vartheta^{(z)}_{\mathbf{m}} = -2\pi\phi y$. Inserting these phases into Eq.~(\ref{weyl:real}) and taking Fourier transform along the $z$ direction, we obtain Eq.~(\ref{Eq: wslabx}).

Since the vector potential breaks the translation symmetry along the $y$ direction, the momentum $k_y$ is no longer a good quantum number, and we are not able to take the Fourier transform in this direction. We can retain the translation symmetry by introducing the so-called magnetic unit cells and the corresponding magnetic Brillouin zone. Nevertheless, keeping the Hamiltonian in the real space representation along the $y$ direction is a simpler choice for our work. The operator can then be written explicitly as
\begin{equation}
	\mathcal{H}_{\text{W}} = \sum_k\begin{pmatrix} \Xi^{\dagger}_{1k}\\ \Xi^{\dagger}_{2k}\\ \vdots\\ \Xi^{\dagger}_{N_xk}\end{pmatrix}^T\begin{pmatrix}
		\Delta & \Omega & 0 & \cdots & 0 \\
		\Omega^{\dagger} & \Delta & \Omega & \cdots & 0\\
		\ddots& \ddots& \ddots& \ddots& \vdots\\
		0 & 0 & \cdots & \Omega^{\dagger} & \Delta
	\end{pmatrix} \begin{pmatrix} \Xi_{1k}\\ \Xi_{2k}\\ \vdots\\ \Xi_{N_xk}\end{pmatrix},
\end{equation}
where $\Xi_{xk}$ are $2N_y$-spinors, and the two $2N_y\times2N_y$ matrices $\Delta$ and $\Omega$ are given by
\begin{align}
	&\Delta = \begin{pmatrix}
		\mathcal{D}_1 & \mathcal{O}_1 & 0 & \cdots & \mathcal{O}^{\dagger}_{N_y}\\
		\mathcal{O}^{\dagger}_1 & \mathcal{D}_2 & \mathcal{O}_2 & \cdots & 0\\
		\ddots& \ddots& \ddots& \ddots& \vdots\\
		\mathcal{O}_{N_y} & 0 & \cdots & \mathcal{O}^{\dagger}_{N_y-1} & \mathcal{D}_{N_y}
	\end{pmatrix}
\end{align}
with $\mathcal{D}_y = [M + \cos(k+2\pi\phi y)]\sigma_z$, $\mathcal{O}_y = (\sigma_z+i\sigma_y)/2$, and $\Omega = \mathbf{1}_{N_y\times N_y}\otimes(\sigma_z-i\sigma_x)/2$. Notice that we keep the periodic boundary condition in the $y$ direction, which requires
$2\pi\phi N_y = 2n\pi$ or $N_y=n/\phi$ with $n = 1,2,\ldots$. On the other hand, as we can write the magnetic flux as a ratio of two integers, i.e., $\phi=p/q$, we choose $n=p$ and thus $q=N_y$ for simplicity. This choice corresponds to a magnetic Brillouin zone that has only one $k$-point in the $y$ direction.
\section{\label{App:Chern}Computing Chern numbers}
To illustrate the quantum Hall effect induced by chiral Landau levels, we compute Hall conductance of the gaps between them using the Streda formula \cite{Streda1982} for a 3D system of sizes $L_x\times L_y\times L_z$. For instance, when the Fermi energy lies in a gap, the Hall conductivity of films perpendicular to the $x$ axis can be determined by
\begin{equation}
	\sigma_{yz} = -e\frac{\partial n(E_F)}{\partial B_x}
\end{equation}
with $n(E_F)$ being the particle density. We transform this formula as
\begin{align}
	&\sigma_{yz} = -e\frac{\Delta n(E_F)}{\Delta B_x} = -\frac{e^2}{h}\frac{\Delta N(E_F)}{L_x\Delta p}\frac{qa_y}{L_y}\frac{a_z}{L_z}.
\end{align}
Here, $N(E_F)$ is the number of states below Fermi energy, $\Phi_{yz}/\Phi_0=p/q$, and the magnetic field is varied by changing $p$ and keeping $q$ constant. The Hall conductance is then given by
\begin{equation}
	G_{yz} = \sigma_{yz}L_x = -\frac{e^2}{h}\frac{\Delta N(E_F)}{\Delta p}\frac{1}{\mathcal{N}_{yz}},
\end{equation}
where $\mathcal{N}_{yz}$ is the number of $\mathbf{k}$-points in the 2D magnetic Brillouin zone. Similarly, for slabs perpendicular to $z$, the Hall conductance reads
\begin{equation}
	G_{xy} = \sigma_{xy}L_z = -\frac{e^2}{h}\frac{\Delta N(E_F)}{\Delta p}\frac{1}{\mathcal{N}_{xy}}.
\end{equation}
However, in order to show a connection with the edge states forming in such QHE, instead of finding the Hall conductance we compute the Chern numbers defined by
\begin{equation}
	\mathcal{C}^{(xy)} = \frac{\Delta N}{\Delta p}\frac{1}{\mathcal{N}_{xy}}\quad\text{and}\quad\mathcal{C}^{(yz)} = \frac{\Delta N}{\Delta p}\frac{1}{\mathcal{N}_{yz}}.
\end{equation}
These Chern numbers give the number of chiral edge modes in our quantum Hall system.
\section{\label{App:confinement}Particle-in-a-box Method}
When solving the slab geometry of a lattice Hamiltonian which does not have surface states, we can obtain its spectrum simply by quantizing the momentum along the confinement axis as in the particle-in-a-box problem.

To illustrate this idea, we consider a particular instance, i.e., the WSM given by Eq.~(\ref{Eq: Weyl}) confined in the $z$ direction with thickness $L_z$. To obtain its spectrum, we often Fourier transform Eq.~(\ref{Eq: Weyl}) along $z$ into real space and get the Hamiltonian
\begin{align}
	&\mathcal{H}_{\text{W}} = \sum_{\mathbf{k},z}\Big\{ d^{\dagger}_{\mathbf{k}z}\big[\left(2-\cos k_x -\cos k_y\right)\sigma_z - \sin k_x\sigma_x\nonumber\\
	&- \sin k_y\sigma_y\big]d_{\mathbf{k}z} - d^{\dagger}_{\mathbf{k}z}\frac{\sigma_z}{2}d_{\mathbf{k}(z+1)} - d^{\dagger}_{\mathbf{k}(z+1)}\frac{\sigma_z}{2}d_{\mathbf{k}z}\Big\}.\label{Eq: compare}
\end{align}
This matrix has a size of $2N_z\times2N_z$ with $N_z$ being the number of lattice sites, $z=1,2,\ldots N_z$. The open boundary condition, or hard-wall boundary condition, is imposed by setting the hopping terms between sites $1$ and $N_z$ to be zero, which implies that the wavefunctions always vanish at sites $0$ and $(N_z+1)$. Hence, the width of our quantum well, or the film thickness, is $L_z=(N_z+1)a_z$. This boundary condition is an appropriate approximation since we are mainly interested in the low-energy limit of our model. Diagonalizing $\mathcal{H}$ gives the energy spectrum of our WSM film.

A more convenient way to get the same result is the aforementioned PiBM, which is to set $k_z\rightarrow\pi\zeta /(N_z+1)$ with $\zeta=1,2,\ldots$ in our lattice Hamiltonian Eq.~(\ref{Eq: Weyl}). Notice that the periodicity of our lattice restricts $0<\zeta\pi/(N_z+1)<\pi$ \cite{Milun2002}, and thus we have the integers $\zeta\in[1,N_z]$. Using this substitution, we can obtain the exact energy spectrum of Hamiltonian Eq.~(\ref{Eq: compare}) just by diagonalizing the matrices
\begin{align}
	&h_{\zeta}(\mathbf{k}) = -\sin k_x\sigma_x - \sin k_y\sigma_y \nonumber\\
	&+ \left[M - \cos k_x - \cos k_y - \cos\left(\frac{\pi\zeta}{N_z+1}\right)\right]\sigma_z.
\end{align}
The eigenvalues are simply
\begin{align}
	\varepsilon_{\zeta}(\mathbf{k}) &= \pm\Bigg[\sin^2k_x + \sin^2k_y \nonumber\\
	+ &\left(M - \cos k_x - \cos k_y - \cos\frac{\pi\zeta}{N_z+1}\right)^2\Bigg]^{\frac{1}{2}}.
\end{align}

Nevertheless, if nontrivial edge states are present, e.g., the WSM confined along $x$, this method becomes a rough approximation as it is unable to demonstrate the localized edge states, but it is still sufficient to demonstrate the formation of bulk subbands due to the quantum confinement.
\section{Semiclassical quantization}
Chiral Landau levels at the two Weyl points are given by
\begin{eqnarray}
	E_c(k_3) &=& \eta t(k_3 - k_W\cos\theta) \nonumber\\
	&&+ t'\left[\frac{1}{2}(k_3 - k_W\cos\theta)^2 + \pi\phi\right]\cos\theta,
\end{eqnarray}
\begin{eqnarray}
	E_c(k_3) &=& -\eta t(k_3 + k_W\cos\theta) \nonumber\\
	&&+ t'\left[\frac{1}{2}(k_3 + k_W\cos\theta)^2 + \pi\phi\right]\cos\theta.
\end{eqnarray}
To obtain the semiclassical quantization of Weyl orbits, we apply the Bohr - Sommerfeld - Onsager quantization condition, i.e., $\oint \mathbf{p}\cdot \mathbf{r} = 2\pi\hbar(n+\gamma)$. We split the line integral into two parts
\begin{itemize}
	\item The integration along the Fermi arcs:
	\begin{equation}
		\int_{\text{arcs}}\mathbf{p}\cdot d\mathbf{r} = \frac{\hbar^2}{eBa^2}S_{\mathbf{k}} \approx \frac{\hbar^2}{e\Phi}k_a2\frac{E}{t} = \frac{\hbar}{\pi\phi}\frac{k_aE}{t},
	\end{equation}
	where $S_{\mathbf{k}}$ is dimensionless, $\Theta$ is the angle determining the confinement direction, and $k_a = 2k_W\sin\Theta$.
	\item The integration along the bulk chiral Landau levels
	\begin{equation}
		\int_{\text{lls}}\mathbf{p}\cdot d\mathbf{r} = \hbar k_32L_z' = 2\hbar L_z'\left(\frac{-t'k_W\sin^2\theta + \sqrt{\Delta}}{t'\cos\theta}\right)
	\end{equation}
	Here, $L_z'=\dfrac{N_z+1}{\cos(\Theta-\theta)}$ is dimensionless, $\theta$ is the angle for the direction of magnetic field, and $\Delta = t^2 - 2t'\cos\theta(\pi t'\phi\cos\theta - E)$.
\end{itemize}
Using these equations, we get the energy levels of Weyl orbits as given in Eq.~(\ref{Eq: minh Weyl orbits}).
\section{\label{App:quantization}Laudau quantization of topological Dirac semimetals}
We consider the continuum model of DSM given by Eq.~(\ref{Eq: Diracconti}). In the presence of a magnetic field $\mathbf{B}\|\hat{z}$, the momenta $(q_x,q_y)$ are replaced by the ladder operators as
\begin{equation}
	q_x\rightarrow \frac{1}{\sqrt{2}l_B}(l^{\dagger}+l),\quad q_y\rightarrow \frac{-i}{\sqrt{2}l_B}(l^{\dagger}-l).\label{eq: ladder1}
\end{equation}
and the Zeeman interaction is added. The Hamiltonian then becomes
\begin{equation}
	H_{\text{D}}(k_z) = \begin{pmatrix}
		H_{\text{D}}^{\uparrow}(k_z) & 0\\ 0 & H_{\text{D}}^{\downarrow}(k_z)
	\end{pmatrix}
\end{equation}
with
\begin{equation}
	\small H_{\text{D}}^{\uparrow}(q_z) = \mathcal{E}_{q_z} + \begin{pmatrix}
		M(q_z) + \dfrac{\mu_Bg_sB}{2} & \dfrac{\sqrt{2}A}{l_B}l^{\dagger}\\
		\frac{\sqrt{2}A}{l_B}l & - M(q_z) + \dfrac{\mu_Bg_pB}{2}
	\end{pmatrix}\nonumber
\end{equation}
and
\begin{equation}
	\small H_{\text{D}}^{\downarrow}(q_z) = \mathcal{E}_{q_z} + \begin{pmatrix}
		M(q_z) - \dfrac{\mu_Bg_sB}{2} & -\dfrac{\sqrt{2}A}{l_B}l\\
		-\dfrac{\sqrt{2}A}{l_B}l^{\dagger} & - M(q_z) - \dfrac{\mu_Bg_pB}{2}
	\end{pmatrix}.\nonumber
\end{equation}
Here, we have $\mathcal{E}(q_z) = C_0 + C_1q^2_z + C_2l_B^{-2}\left(2l^{\dagger}l + 1\right)$ and $M(q_z) = M_0 + M_1q_z^2 + M_2l_B^{-2}\left(2l^{\dagger}l + 1\right)$. According to Ref. \cite{Cano2017}, the parameters are $A = 0.889$~eV{\AA}, $M_0=-0.0205$~eV, $M_1=18.77$~eV{\AA}$^2$, $M_2=13.5$~eV{\AA}$^2$, $C_0=-0.0145$~eV, $C_1=10.59$~eV{\AA}$^2$, and $C_2=11.5$~eV{\AA}$^2$. A numerical diagonalization of $H_{\text{D}}(k_z)$ yields the spectrum in Fig.~\hyperref[fig:7]{7(a)}. Besides, using the eigenvectors $(\ket{0},0,0,0)^T$ and $(0,0,0,\ket{0})^T$, we can analytically obtain the chiral Landau bands
\begin{align}
	\epsilon_{0\uparrow}(q_z) = C_0+M_0 &+ (C_1+M_1)q_z^2 \nonumber\\
	&+ (C_2+M_2)\frac{eB}{\hbar} + \frac{\mu_Bg_sB}{2}
\end{align}
and
\begin{align}
	\epsilon_{0\downarrow}(q_z) = C_0-M_0 &+ (C_1-M_1)q_z^2 \nonumber\\
	&+ (C_2-M_2)\frac{eB}{\hbar} - \frac{\mu_Bg_pB}{2}.
\end{align}

On the other hand, the hyperbolic model of Cd$_3$As$_3$ proposed by Ref.~\cite{Cano2017} can be obtained simply by replacing the function $\tilde{\mathcal{M}}_(\mathbf{q})$ with
\begin{equation}
	\tilde{\mathcal{M}}^{(j)}(\mathbf{q}) = M_0^{(j)} + \sqrt{(M^{(j)}_3)^2 + (M^{(j)}_1q_z)^2} + M_2^{(j)}(q_x^2 + q_y^2).
\end{equation}
Here, the parameters are $A^{(j)} = 2.75$~eV{\AA}, $M_0^{(j)}=-0.06$~eV, $M_1^{(j)}=96$~eV{\AA}$^2$, $M_2^{(j)}=18$~eV{\AA}$^2$, $M_3^{(j)}=0.05$~eV{\AA}$^2$, $C_0^{(j)}=-0.219$~eV, $C_1^{(j)}=-30$~eV{\AA}$^2$, $C_2^{(j)}=-16$~eV{\AA}$^2$. Following the same calculation as before, we get the spectrum in Fig.~\hyperref[fig:7]{7(b)} and the analytical expressions
\begin{align}
	\epsilon_{0\uparrow}(q_z) = C_0+\tilde{M}_0 &+ C_1q_z^2 + \sqrt{M_3^2 + \tilde{M}_1k_z^2}\nonumber\\
	&+ (C_2+M_2)\frac{eB}{\hbar} + \frac{\mu_Bg_sB}{2},\\
	\epsilon_{0\downarrow}(q_z) = C_0-\tilde{M}_0 &+ C_1q_z^2 - \sqrt{M_3^2 + \tilde{M}_1k_z^2} \nonumber\\
	&+ (C_2-M_2)\frac{eB}{\hbar} - \frac{\mu_Bg_pB}{2}.
\end{align}

\bibliography{Minh2021}

\end{document}